\documentclass{aa}
\usepackage{natbib, twoopt}
\usepackage{newtxtext,newtxmath}
\usepackage[T1]{fontenc}
\usepackage[normalem]{ulem}
\usepackage{acronym}

\usepackage{graphicx}   
\usepackage{amsmath}    
\usepackage{amssymb}    
\usepackage[usenames,dvipsnames]{xcolor}
\usepackage{soul}
\usepackage{enumitem} 
\usepackage[breaklinks=true]{hyperref}
\bibpunct{(}{)}{;}{a}{}{,}

\newcommand{\kms}{\,{\rm km\,s^{-1}}}
\newcommand{\msun}{\,{\rm M_\odot}}
\newcommand{\accunit}{\,{\rm km^2\,s^{-2}\,kpc^{-1}}}
\newcommand{\kpc}{\,{\rm kpc}}
\newcommand{\pc}{\,{\rm pc}}
\newcommand{\Gyr}{\,{\rm Gyr}}
\newcommand{\Myr}{\,{\rm Myr}}

\newcommand{\update}[1]{\textcolor{black}{#1}}

\title{ Back to the present: A general treatment for the tidal field from the wake of dynamical friction}
\author{
Rain Kipper\inst{1}\thanks{E-mail: rain.kipper@ut.ee} \and
Peeter Tenjes\inst{1}\and 
Mar\'ia Benito\inst{1}\and
Punyakoti Ganeshaiah Veena\inst{2}\and
Aikaterini Niovi Triantafyllaki\inst{1}\and
Indrek Vurm\inst{1}\and
Moorits Mihkel Muru\inst{1}\and
Maret Einasto\inst{1}\and
Elmo Tempel\inst{1,3}
}
\institute{Tartu Observatory, University of Tartu, Observatooriumi 1, T\~oravere 61602, Estonia
\and Department of Physics and the Asher Space Science Institute, Technion – Israel Institute of Technology, Haifa 3200003, Israel
\and Estonian Academy of Sciences, Kohtu 6, 10130 Tallinn, Estonia
}

\date{Received 20 June 2023 / Accepted 10 October 2023}

\begin{document}

\label{firstpage}
\abstract{
Dynamical friction can be a valuable tool for inferring dark matter properties that are difficult to constrain by other methods. Most applications of dynamical friction calculations are concerned with the long-term angular momentum loss and orbital decay of the perturber within its host. This, however, assumes knowledge of the unknown initial conditions of the system.
} 
{
We advance an alternative methodology to infer the host properties from the perturber's shape distortions induced by the tides of the wake of dynamical friction, which we refer to as the tidal dynamical friction.
}
{
As the shape distortions rely on the tidal field that has a predominantly local origin, we present a strategy to find the local wake by integrating the stellar orbits back in time along with the perturber, then removing the perturber's potential and re-integrating them back to the present. This provides perturbed and unperturbed coordinates and hence a change in coordinates, density, and acceleration fields, which yields the back-reaction experienced by the perturber. }
{
The method successfully recovers the tidal field of the wake based on a comparison with N-body simulations.
We show that similar to the tidal field itself, the noise and randomness of the dynamical friction force due to the finite number of stars is also dominated by regions close to the perturber. Stars near the perturber influence it more but are smaller in number, causing a high variance in the acceleration field. These fluctuations are intrinsic to dynamical friction. We show that a stellar density of $0.0014\msun\,{\rm kpc^{-3}}$ yields an inherent variance of 10\% to the dynamical friction. 
}
{
The current method extends the family of dynamical friction methods that allow for the inference of host properties from tidal forces of the wake. It can be applied to specific galaxies, such as Magellanic Clouds, with Gaia data. 
}


\maketitle


\section{Introduction}\label{sec:introduction}
A perturber moving through a host -- an environment of collisionless particles -- is subject to dynamical friction.  The perturber distorts the host's density distribution, and this distortion alters the velocity of the perturber. Dynamical friction was first described by \citet{Chandrasekhar:1943} for an infinite, homogeneous, and isotropic environment and later advanced by \citet{Tremaine:1984} for a spherically symmetrical host. Since then, several other generalisations have been developed, for example, multi-component systems and ellipsoidal velocity distribution \citep{Bonetti:2021, Moreno:2022}. \citet{Heyvaerts:2017} evaluated dynamical friction via diffusion coefficients and showed equivalence of many representations. \citet{Banik_2021b} reformulated the dynamical friction modelling in an orbital-based framework with the aim of describing the central regions of galaxies. The response of a galaxy to a massive perturber has been studied with simulations \citep[see e.g.][]{tangoforthree, Vasiliev:2023} as well as semi-analytically \citep{Bonetti:2021}. Most dynamical friction studies emphasise the loss of integrals of motion of the perturber or its influence on the host. The aim of this paper drifts from these goals, as we aim to infer the host properties based on the distorting forces induced by the wake of the dynamical friction \update{towards} the perturber. 

Dynamical friction depends on the velocity distribution of a host's particles, and one can use this dependency to infer the host's dark matter (DM) kinematics. Since the counterpart particle of DM has not yet been found, an enormous effort is being made to determine its nature from astronomical observations. Dark matter inference studies on large and galactic scales have limitations due to the precision of observations \citep{Fisher:2022} and/or the modelling methods \citep{Roper:2022, Downing:2023}. On smaller scales, a particularly powerful discriminator of the DM nature is the mass function of the DM substructures \citep{2020PhRvD.101j3023B, Lovell_2021, Nadler_2021}. Dynamical friction offers an alternative way to study the small-scale effects induced by hypothetical microphysical properties of DM \citep{Hui_2017, Hartman:2021}. Moreover, its applications are much broader than DM inference, as dynamical friction is the underlying phenomenon that influences a wide variety of astrophysical systems, from stellar streams and globular clusters to the orbits of satellite galaxies and galaxy merging times \citep{Kipper_stream, Kipper_lopsided, Pflamm23, CorreaMagnus:2022}.

In essence, dynamical friction is a process that makes a heavier object slow down in comparison to its environment. As the slowing down effect depends on the host's environment, dynamical friction can also be used as a tool to study both host and perturber properties. To put this tool into action, we distinguished two observational regimes.

\update{In the first regime, the inference can be based on a long-term influence on the perturber orbit.} Weak and persistent dynamical friction influences an object over an extended period of time, thus altering the radial location of the object in a galaxy or cluster. A prominent example of a long-term effect is core stalling. The globular clusters in the Fornax dwarf are an example of this, as they have not yet fallen to the centre (see \citet{Meadows_2019, Bose_2019}). Another example of long-term dynamical friction manifestation is its influence on the relative position of a globular cluster compared to a stream centre \citep{Kipper_stream}. 

\update{The inference can be based on the tidal field by the wake. } Dynamical friction decreases the velocity of the perturber and additionally causes tidal forces to affect the perturber. Some parts of the perturber slow down faster than others, causing distortions in shape. This was first studied by \citet{Mulder:1983} and in the context of inferring DM properties by \citet{Garavito_Camargo_2019} and \citet{Kipper_lopsided}. \citet{Kipper_lopsided} studied an isolated galaxy moving in a void environment and showed that dynamical friction due to intergalactic DM can produce lopsided gas and stellar distributions. 
In this paper, our attention is focused on the tidal aspect of the dynamical friction. We develop a model with the emphasis to use as few simplifying assumptions as possible when using the method. 

Simplifying assumptions and their validity are central for most dynamical methods. Solving the equations of motion that include dynamical friction requires using simplifying assumptions that might influence the accuracy of the results. For example, the classical methods of Chandrasekhar and \citet{Tremaine:1984} assume that the underlying medium is uniform, spherical, or axisymmetric. In many galaxies, this is not the case. For example, the Milky Way disc has perturbations and non-equilibrium features \citep{Antoja:2018}. Another common assumption is to use nearly circular orbits for the perturber. However, most globular clusters do not have circular orbits \citep{Vasiliev:2021}. Lastly, several studies assume that the density distribution of a galaxy is time independent and thus discard secular evolution of the host \citep{Chandrasekhar:1943, Tremaine:1984, Banik_2021, Desjacques_2022}, but some evolution has been shown to occur \citep{Gunawardhana:2013, Kipper_galevol, Hashemizadeh:2022}. These kinds of assumptions lead to biases in dynamical friction calculations. Nonetheless, relaxing these assumptions poses problems for analytical calculations. Analytical description of the model is close to mandatory in case one wishes to infer something about specific objects using likelihood, although there are a few exceptions, such as \citet{tangoforthree, Reza:2022}. Moreover, numerical simulations can be used for calculating dynamical friction, but they are time-consuming, and since slight deviations from initial conditions may give quite different results \citep{Genel_2019}, they can only provide a statistical understanding \citep{2017MNRAS.464.2882A}. 

The tidal aspect of dynamical friction is not well studied. The pioneering paper by \citet{Mulder:1983} solved it in the case of an isotropic velocity distribution and concluded that tidal effects are mainly from regions near the perturber. 
Another paper discussing tidal aspects of dynamical friction is \citet{Kipper_lopsided}, where we showed that isolated galaxies can form lopsidedness without mergers just by moving through large-scale structures. 

In this paper, we propose a semi-analytic method to calculate dynamical friction that relaxes many of the commonly adopted assumptions. This enables a more accurate estimation of the tides of dynamical friction and the sensitivity needed to investigate the nature of DM.  
The method can be used with real Gaia data to study the Milky Way. Testing of the method indicates that the process of the dynamical friction is subject to fluctuations  of forces from background stars. This aspect has not obtained sufficient attention yet. 

The paper is organised as follows. Section~\ref{sec:method} presents our method for calculating dynamical friction and its extension for tidal field evaluations. Section~\ref{sec:simulation} summarises the simulated tools that are used in Section~\ref{sec:validation} to validate the method and evaluate its limitations. Then, Section~\ref{sec:sampling_noise} assesses the dependence of the dynamical friction estimates on sampling noise. We conclude with a discussion and a summary in Sections~\ref{sec:discussion} and \ref{sec:conclusion}.

\section{Method}\label{sec:method} 

From this point and onwards, we refer to the tides from the wake of dynamical friction as `tidal dynamical friction'. Preserving the complexity of phase-space density with maximal accuracy is important for realistic tidal dynamical friction estimates. For example, \citet{Petts:2016} and \citet{ Leaman:2021} showed that deviations from the simple Maxwellian velocity distribution significantly affect dynamical friction estimates. Hence, we must be able to accurately account for all the complexities. We describe the host as a set of particles, each with its own position and velocity, which allows for an accurate characterisation of the phase-space density. However, this particle-based description is challenging to acquire observationally, which may necessitate the use of smooth distribution functions. In this work, we develop a framework for tidal dynamical friction calculations suitable with both phase-space and particle-based descriptions of the system. 

Our aim is to obtain the tidal dynamical friction estimation, which is mostly determined by regions in the vicinity of the perturber \citep{Mulder:1983}. We kept this feature in mind when deriving the methodology. 

\subsection{Basic framework}\label{sec:basic_framework}
The setup for estimating dynamical friction consists of an environment -- the host -- and a massive perturber moving through the host. The acceleration field caused by dynamical friction is the difference between the acceleration fields generated by the host's density distribution when the perturber is present or absent. Our definition of dynamical friction is slightly wider than what is commonly used, as it includes evaluation points other than the perturber itself. We denote the position of a particle in the absence of the perturber as ${\bf x}_{0}$, and in the presence of the perturber, we denote it as ${\bf x}_{1}$. The total number of particles in the system is $N_{\rm tot}$. We describe the accelerations at an arbitrary point ${\bf x}$ in the host as ${\bf a}_{0}$ in the absence of the perturber and as ${\bf a}_{1}$when the perturber exists: 
\begin{eqnarray}
    {\bf a}_{0}({\bf x}, t) &=& \sum_{i=1, i\ne j}^{N_{\rm tot}} K_{0} \frac{{\bf x}_{0,i} - {\bf x}}{|{\bf x}_{0,i} - {\bf x}|}   \label{eq:anp}\\
    {\bf a}_{1}({\bf x}, t) &=& \sum_{i=1, i\ne j}^{N_{\rm tot}} K_{1}\frac{{\bf x}_{1,i} - {\bf x}}{|{\bf x}_{1,i} - {\bf x}|}  \label{eq:ap}, 
\end{eqnarray}
respectively. The condition $i\ne j$ is only used when the acceleration is evaluated for particle $j$. The term $K$ denotes a kernel describing a particle's acceleration, which can take different forms depending on the system we are interested in. For example, in simulation comparison, the kernel should include softening. For point sources with mass $m$, the kernel is Newton's gravitational acceleration:
\begin{eqnarray}
    K &=& \frac{Gm}{r^2}, 
    \label{eq:kernel}
\end{eqnarray}
where $r$ in $K_{0}$ corresponds to $|{\bf x}_{0,i} - {\bf x}|$ for the unperturbed state, while $r$ in $K_{1}$ corresponds to $|{\bf x}_{1,i} - {\bf x}|$ for the perturbed state.

These equations hold at any fixed time. Dynamical friction manifests as the difference between these acceleration fields:
\begin{eqnarray} \label{eq:acc_DF}
    {\bf a}_{\rm DF} \equiv {\bf a}_{1} - {\bf a}_{0}\label{eq:adf}.
\end{eqnarray}
We call the ${\bf a}_{\rm DF}$ total `one' when summing in Eqs.~\eqref{eq:anp} and \eqref{eq:ap} is done over all the points and refer to the total as `local' when only points in the vicinity of the perturber are included. In the first case, ${\bf a}_0$ and ${\bf a}_1$ correspond to the acceleration field of the host $-\nabla \Phi_{\rm h}^0 ({\bf x}, t) = {\bf a}_0$ or $ -\nabla \Phi_{\rm h}^1 ({\bf x}, t) = {\bf a}_1$ in the perturbed state. 
The difficulty regarding dynamical friction calculations is that, generally, only one of these acceleration fields is known, either~\eqref{eq:anp} or~\eqref{eq:ap}. In order to get one from the other, we devised a method that is based on initially integrating the particles and perturber back in time until the perturber has a negligible influence on the particles. Then, we remove or add the perturber's potential to the host and re-integrate the orbits back to the present. A similar method was developed by \citet{CorreaMagnus:2022} to evaluate the mass of the Milky Way. Although from its theoretical aspects, this is a simple procedure, from the practical aspects, such as selection function, uncertainties, and the accuracy of gravitational potential, this method poses several questions during its implementation.

We define the time $t=0$ as the moment at which we wish to calculate the tidal dynamical friction and assume that we know the unperturbed distribution of particles at that time, that is, their positions ${\bf x}_0(t=0)$ and velocities ${\bf v}_0(t=0)$. The equations of motion describe the movement of particles, where the host potential fully determines the acceleration of these particles:
\begin{eqnarray}
    \frac{{\rm d}{\bf v}_0}{{\rm d}t} = -\nabla\Phi_{\rm h}^0({\bf x}_0, t).\label{eq:dotv0}
\end{eqnarray} 
The host potential depends on position and time. The explicit time dependence can include contributions from a rotating bar or other external processes (i.e. non-stationarity). However, in the presence of the perturber, described with potential $\Phi_{\rm p}$ and coordinate ${\bf x}_{\rm p}$, the particle accelerates according to 
\begin{eqnarray}
    \frac{{\rm d}{\bf v}_1}{{\rm d}t}  = -\nabla {\Phi}_{\rm h}^{1}({\bf x}_1, t) - \nabla\Phi_{\rm p}({\bf x}_1 | {\bf x}_{\rm p}(t), t)\label{eq:dotv1}.
\end{eqnarray}
Unfortunately, regarding the use of Eq.~\eqref{eq:dotv1} to calculate the orbit of a disturbed particle, we do not yet know the potential $\Phi^1$. Thus, we calculate it iteratively and replace $\Phi^1$ with a guessed potential  $\Phi^{1*}$, namely,
\begin{eqnarray}
    \frac{{\rm d}{\bf v}_1}{{\rm d}t}  = -\nabla {\Phi}_{\rm h}^{1*}({\bf x}_1, t) - \nabla\Phi_{\rm p}({\bf x}_1 | {\bf x}_{\rm p}(t), t) \label{eq:dotv11}.
\end{eqnarray}
For the first iteration we take
\begin{eqnarray}
\Phi^{1*} = \Phi^0.\label{eq:first_iteration}
\end{eqnarray}
Taking this approach means we approximate the potentials without taking into account internal deformations of the galaxies, that is,  for the first iteration, the self-gravity of the wake is not accounted for. This regime is satisfied when two galaxies gravitate towards each other, and the smaller has undergone only one pericentre passage 
\citet{Vasiliev:2022, CorreaMagnus:2022}.
For subsequent iterations, we use the following approximation to $\Phi^{1*}$
\begin{eqnarray}
    \nabla \Phi^{1*} = \nabla \Phi^0 + {\bf a}_{\rm DF*},
\end{eqnarray}
where ${\bf a}_{\rm DF*}$ is evaluated\footnote{The details of orbit integration and whether ${\bf a}_{\rm DF}^*$ is approximated or used with direct summing are not part of the basic framework and are specific for each implementation.} using Eq.~\eqref{eq:ap}, but the coordinates we use are ${\bf x}_1^*$ instead of ${\bf x}_1$. 
Upon several iterations, $\Phi^{1*}$ approaches $\Phi^1$. 
The perturber's trajectory ${\bf x}_{\rm p}(t)$ should also be calculated according to
\begin{eqnarray}
    \frac{{\rm d}{\bf v}_{\rm p}}{{\rm d}t} = -\nabla\Phi^{1*}. \label{eq:dotv1_*}
\end{eqnarray}
We assume that the initial conditions of the perturber (coordinates ${\bf x}_{\rm p}$ and velocity ${\bf v}_{\rm p}$) are known. 

Solving Equations \eqref{eq:dotv0} or \eqref{eq:dotv1} numerically requires knowing the initial conditions or the conditions at time $t=0$. In the case where the perturber is observed, the initial conditions are ${\bf x}_{1,i}(t=0) = {\bf x}_{\rm obs,i}$. If we want to calculate the dynamical friction of a hypothetical perturber located at ${\bf x}_{\rm p}$, then the initial conditions are ${\bf x}_{0, i}(t=0) = {\bf x}_{\rm obs, i}$. From now on, we assume the latter case, but we can derive the equations for the former case in a similar manner.

\begin{figure}
    \centering
    \includegraphics{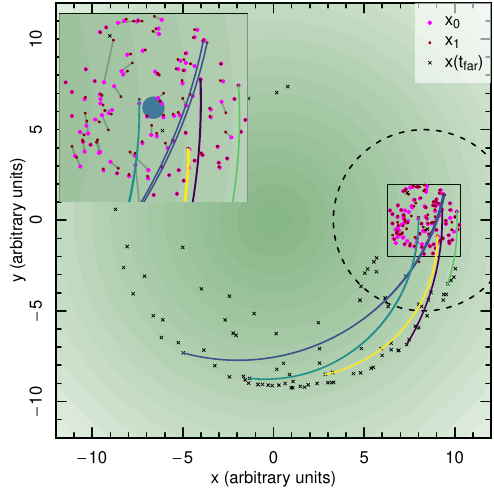}
    \caption{Illustration of the dynamical friction framework. The magenta points describe the points of the host in the vicinity of a perturber (depicted as a blue dot in the enlarged sub-plot). Different coloured lines show a small sub-sample of integrated orbits, both backwards and forwards in time. The dashed circle shows the distance from the perturber where the $t_{\rm far}$ is reached. The acceleration field difference caused by unperturbed (${\bf x}_0$) and perturbed (${\bf x}_1$) points  gives the dynamical friction. In the enlarged image, we connect the corresponding points with a grey line.}
    \label{fig:illustration_bttf}
\end{figure}

In order to convert the position of a particle from the unperturbed case (${\bf x}_0$) to the perturbed one (${\bf x}_1$), we make the assumption that there is a time $t_{\rm far}\le0$ where the particle on its orbit is so far from the perturber that the perturber's contribution to the particle acceleration is negligible. Mathematically, this is equivalent to the following condition:
\begin{eqnarray}
    |\nabla\Phi_{\rm h}({\bf x}_0(t_{\rm far}))|
    \gg
    |\nabla\Phi_{\rm p}({\bf x}_0(t_{\rm far})|.
    \label{eq:main_assumption}
\end{eqnarray}
The practical limitations of this condition to exist are explained in Sect.~\ref{subsec:turnaround}. At this point, we may integrate the orbit for a particle $i$ at the acceleration field ${\bf a}_0 ({\bf x}, t)$ with the initial condition ${\bf x}_{0, i}(t=0) = {\bf x}_{\rm obs, i}$ until $t_{\rm far}$ and derive ${\bf x}_{0,i}(t_{\rm far})$. At $t=t_{\rm far}$ the particle has no information on whether the perturber exists or not. 
Therefore, at $t=t_{\rm far}$,
\begin{eqnarray}
    {\bf x}_1(t_{\rm far}) = {\bf x}_0(t_{\rm far})\label{eq:ic_in_past},
\end{eqnarray}
and we may take ${\bf x}_{1*,i} (t_{\rm far})$ as the initial condition to integrate the orbit ${\bf x}_{1*,i}$ until $t=0$. Again, the asterisk denotes the potential approximation $\Phi^{1*}$ in use. With this procedure, we are able to calculate ${\bf x}_{1*}(t=0)$ from ${\bf x}_0(t=0)$ for all the particles. 
We then enter the newly acquired positions ${\bf x}_{1*}$ into Eq.~\eqref{eq:ap}, which are the basis to calculate (tidal) dynamical friction with Eq.~\eqref{eq:adf}. This method is illustrated in Fig.~\ref{fig:illustration_bttf}.

The equations above correspond to the dynamical friction for a particle-based description. In the case of a continuous description of the system, the equations change slightly:
\begin{eqnarray}
    {\bf a}_0({\bf x}) = \int K_{0} \frac{{\bf x}_0 - {\bf x}}{|{\bf x}_0 - {\bf x}|}f({\bf x}_0,{\bf v}_0){\rm d}{\bf x}_0{\rm d}{\bf v}_0 ,\label{eq:analy_nonpert}\\
    {\bf a}_1({\bf x}) = \int K_{1} \frac{{\bf x}_1 - {\bf x}}{|{\bf x}_1 - {\bf x}|}f({\bf x}_1,{\bf v}_1){\rm d}{\bf x}_1{\rm d}{\bf v}_1  \label{eq:analy_pert}.
\end{eqnarray}
We denote $f$ as the phase-space distribution normalised so that integration over velocities is the ordinary spatial density: $\int f {\rm d}{\bf v}=\rho$.

\subsection{The tidal aspect of dynamical friction}
\label{sec:tidaleffects}
The last section describes the dynamical friction in a broad context. In this section, we shift the attention to tidal aspects that originate locally. As local regions are near the perturber and they are able to provide high-leverage contributions, noise, and fluctuations, we can estimate how much the sampling noise propagates to dynamical friction. 

To gain insights into the forming of dynamical friction and its fluctuations, we return to Eqs.~\eqref{eq:anp}, \eqref{eq:ap}, and \eqref{eq:adf}.  Rewriting \eqref{eq:adf} in a simplified, one-dimensional form for a single particle contribution, we obtain
\begin{eqnarray}
    \delta\!a = \frac{Gm}{r^2} - \frac{Gm}{(r-\delta x)^2} = 
    Gm\left[ \frac{\delta\!x^2 - 2r\delta\!x}{r^2(r^2 - 2r\delta\!x + \delta\!x^2)} \right].
\end{eqnarray}
We use $r$ to denote the distance of the unperturbed position to the evaluated point, and $\delta\!x = x_1 - x_0$ is the shift between perturbed and unperturbed positions (i.e. the source of dynamical friction). In case the force-inducing pair is far from the evaluated point ($\delta\!x \ll r$), the above relation becomes
\begin{eqnarray}
    \delta\!a = \frac{-2Gm\delta\!x}{r^3}.
    \label{eq:daofaparticle}
\end{eqnarray} 
That is, the dynamical friction asymptotically has a tidal-like ($r^{-3}$) behaviour. If the force-inducing pair is acting as the source of the tidal force (i.e. different distances ($\delta\!y$) are affected differently by this pair), then the contribution analogously becomes\begin{eqnarray}
    \delta\!a_{\rm tide} = \frac{-2Gm\delta\!x}{r^3} - \frac{-2Gm\delta\!x}{(r-\delta\!y)^3} \propto r^{-4}.\label{eq:r4}
\end{eqnarray}
We use $\delta\!a_{\rm tide}$ in order to denote the contribution of the pair to the tidal dynamical friction.

The dynamical friction does not result from a single pair of points but from all points in the host. We divided the points among shells with a mean distance to the perturber denoted as $r$. Combining the shell contributions into the total signal is done repeatedly (for Newton acceleration, dynamical friction, and tidal dynamical friction); hence, we denote the total signal as $F$ and its variance as $\sigma^2$, the kernel (gravitational force $\propto r^{-2}$, dynamical friction $\propto r^{-3}$, or tidal force from dynamical friction $\propto r^{-4}$) as $w$, and the contribution from each shell $i$ as $W_i$. 

Each shell's contribution $W_i$ is proportional to the number of points in the shell and the kernel $\propto N_{\rm pt} w$. 
We considered a shell with thickness $\Delta r$. The number of particles \update{in the shell} is proportional to the density ($n$) and the volume of the shell: $W \propto n r^2 \Delta r w$. The corresponding uncertainty for each shell originates from sampling, as the Poisson distribution gives contributions only from the density part, while the kernel is constant $\Delta W_i \propto \sqrt{N_{\rm pt}}w \propto \sqrt{r^2\,\Delta r}w = r w \sqrt{\Delta r}$ 
and variance $\Delta W_i^2 \propto w^2  r^2 \Delta r$. \update{After finding each shell contribution, they need to be combined.} Summing or integrating the contributions over all shells at the thin shell limit ($\Delta r \rightarrow {\rm d}r$) to either signal ($W$) or its variance ($\Delta W$) gives $F \propto \sum W_i = \int r^2 w {\rm d}r$ or $\sigma^2 \propto \sum (\Delta W_i)^2 = \int r^2 w^2 {\rm d}r$. 
In order to find the signal-to-noise ratio, we divided the two $\sqrt{\sigma^2}/F$. 

Next, we compared the convergence of signal and noise in cases of Newton gravity, dynamical friction, and the tidal dynamical friction in the context of signal, noise, and signal-to-noise ratio (see \update{ Table~}\ref{tab:different_signals}).
\begin{table}
\caption{\update{Contribution from all shells to different quantities. }}\label{tab:different_signals}
\begin{tabular}{ ll ll}
\hline
     $w$ &  $F$ & $\sigma$ & $\sigma/F$\\
     \hline
     Newtons gravity $\propto r^{-2}$ & $\propto r^{}$ & $\propto r^{-1}$ & $\propto r^{-1.5}$ \\
     Dynamical friction $\propto r^{-3}$ & $\propto \ln r$ & $\propto r^{-1.5}$ & $\propto r^{-1.5}/\ln r$ \\
     Tidal dyn. friction $\propto r^{-4}$ & $\propto r^{-1}$ & $\propto r^{-2.5}$ & $\propto r^{-1.5}$\\
     \hline
\end{tabular}
\end{table}
As can be seen in the case of dynamical friction, the forces accumulate proportionally with $\ln r$, precisely as in case of Chandrasekhar dynamical friction. But a new result is the formation of the signal-to-noise ratio. In the case of dynamical friction, the ratio accumulates faster, indicating that the inner parts contribute more to noise accumulation than in the case of Newton's law of gravity. For the tidal dynamical friction, one can see that the total contribution behaves as $r^{-1}$, causing inner parts to be far more relevant in the signal accumulation than the outer parts. This was reported in \citet{Mulder:1983}, and we reproduce it in this work. Analogously, one can also see that the noise originates from the vicinity of the perturber. 

The previous line of reasoning relies on the assumption of a homogeneous environment, but this is not generally the case, and this analysis only provides an understanding of what to expect. For the general uncertainty estimations from sampling noise, we used bootstrapping \citep{bootstrap_mainpaper}. 

The tidal dynamical friction provides the force distorting the perturber. How exactly the perturber is distorted is shown by the shapes of the orbits bound to the perturber. Denoting $\delta{\bf x}_{\rm tidal}$ as the distance from the perturber where we want to evaluate a particle orbit, the tidal dynamical friction term 
\begin{eqnarray}
    \delta\!{\bf a}_{\rm DF} = 
    {\frac{{\rm d}{\bf a}_{\rm DF}}{{\rm d}{\bf x}} }\Bigr|_{{\bf x} = {\bf x}_{\rm p}} \delta\!{\bf x}_{\rm tidal}
    \label{eq:theory_tidal}
\end{eqnarray}
must be included in the acceleration field. We take the derivative at the location of the perturber. The tidal field from the dynamical friction affects the orbits in or around the perturber, and they are evaluated based on acceleration
\begin{eqnarray}
    \dot{{\bf v}} - \dot{{\bf v}}_{\rm p}=-\nabla\Phi_{\rm p}+\delta\!{\bf a}_{\rm DF} - \nabla(\Phi_{\rm h}^0 - \Phi_{\rm h}^0|_{{\bf x}_{\rm p}}). 
\end{eqnarray}
The first term describes the perturber's potential, the second describes the tidal dynamical friction, and the third is the host's tidal field.

To summarise, the current method is aimed at inferring the host properties using the shape distortions the wake induces to the perturber. The shape distortions originate from the dynamical friction differences at points in the vicinity of the perturber. The mean dynamical friction averaged over all points near the perturber does not contribute to the distortions but causes the slowing down of the perturber. In this way, we can split the dynamical friction into two parts:
\begin{eqnarray}
    {\bf a}_{\rm DF} = \bar{{\bf a}}_{\rm DF} + {\bf a}_{\rm DF}^\star,
\end{eqnarray}
where $\bar{{\bf a}}_{\rm DF} \equiv \langle {\bf a}_{\rm DF} \rangle$ describes the average pull of the perturber and ${\bf a}_{\rm DF}^\star$ describes the differences from the average. The mean value $\bar{{\bf a}}_{\rm DF}$ originates from quite different regions in the host due to its $\propto \ln r$ dependence, but the tidal field caused by the wake ($\nabla {\bf a}_{\rm DF}$) originates mainly from the vicinity of the perturber since contributions from different distances are accumulated with weights $\propto r^{-1}$. Therefore, in order to recover the total dynamical friction, we need to recover the distortions all around the galaxy, while tidal field recovery is local. In the current approach, we are interested in tidal fields, hence ${\bf a}_{\rm DF}^\star$, and we may therefore evaluate dynamical friction only within a finite region. 

\subsection{Determining the extent of orbit integration}
\label{subsec:turnaround}
To use our dynamical friction framework, one needs to use the condition \eqref{eq:main_assumption} to decide the time span of orbit integration or $t_{\rm far}$. In this section, we examine the basis to find $t_{\rm far}$ in more detail.

We assume that in any practical application, we do not know the gravitational potential fully and have to rely on some approximation for it. The main ones in the literature are axial or spherical symmetry and equilibrium, but we are also concerned with more fine-grained mismatches. By neglecting asymmetries and the time evolution of the acceleration field in our dynamical friction implementation, we biased the estimation via orbit integration of the perturbed positions $\textbf{x}_1$. The mismatch between $\textbf{x}_{\rm 1, recovered}$ and $\textbf{x}_1$ scales with the integration time, and an upper limit is $x_{\rm mismatch}=a_{\rm \sigma} t_{\rm far}^2/2$  in case of a single larger fluctuation or a random walk for many smaller mismatches. We denote the fluctuation levels of the acceleration field as $a_{\rm \sigma}$. We should minimise $x_{\rm mismatch}$. That is, we needed to find the minimum value of $t_{\rm far}$ (or the maximum value of $a_{\rm far}$) that satisfies condition~\eqref{eq:main_assumption}. 
Further, we required that the shift in the position of a star due to the perturber (the dynamical friction) be greater than $x_{\rm mismatch}$; otherwise, the change in perturbed and unperturbed positions is an artefact of the modelling or implementation assumptions  that enter the acceleration description. As $x_{1} - x_{0} \gtrsim a_{\rm far} t_{\rm far}^2/2$, $a_{\rm far} > a_\sigma$\update{, the condition}~\eqref{eq:main_assumption} can be rewritten as
\begin{equation}
    |\nabla\Phi_{\rm h}({\bf x}_0(t_{\rm far}))|
    \gg
    |\nabla\Phi_{\rm p}({\bf x}_0(t_{\rm far})| > |\nabla\Phi_{\rm h, fluctuations}|.
    \label{eq:practical_main_assumption}
\end{equation}
In this way, we minimised mismatches in the positions of stars due to irregularities in the acceleration field of the host galaxy not included in our smooth description that enters the orbit integration in Eqs.~\eqref{eq:dotv0} and \eqref{eq:dotv1}. We write the above condition as
\begin{eqnarray}
a_{\rm far} = N a_\sigma, \label{eq:N_definition}
\end{eqnarray}
where $N$ is the free parameter of the modelling, and it depends on the noise level of the gravitational potential of the host ($a_\sigma$). 

\subsection{An alternative orbit integration scheme}\label{sec:adiabatic_introduction_theory}
To evaluate the tidal dynamical friction with the current method, the condition \eqref{eq:main_assumption} or \eqref{eq:practical_main_assumption} must be fulfilled. For stars that are near the host's centre or have some specific co-movement with the perturber, the condition can never be fulfilled or is always fulfilled. Hence, we needed an alternative for these cases. 

To overcome this issue, we modified how these conditions can be fulfilled in a similar manner to \citet{Tremaine:1984}. These authors introduced the perturber adiabatically to the system by including the term $\exp(\eta t)$ in the density of the perturber with the $\eta$ value being very small in order to mimic the buildup of the system. That way, the perturber starts with almost zero initial mass when $t\ll 0$ and develops gradually to its final mass when time equals zero. We applied a similar recipe to the acceleration field of the perturber by replacing it with
\begin{eqnarray}
    \nabla \Phi_{\rm pert} \rightarrow \tau(t)\nabla \Phi_{\rm pert}.
\end{eqnarray}
The $\tau(t)$ describes multiplicative correction to the acceleration (or, equivalently, density). In this case, the conditions \eqref{eq:main_assumption} or \eqref{eq:practical_main_assumption} are fulfilled using the adiabatic introduction, not from physical separation. This fulfilment can be achieved by taking $t_{\rm far}$ as a fixed time when the perturber had a certain, sufficiently small mass or by retaining individual values of $t_{\rm far}$ for each particle. The latter possibility does not solve the issue when the conditions are always fulfilled nor when many of the very minor contributions to the acceleration field are included.  

This modification poses some changes to the calculation's algorithm, as it is necessary to specify the form of $\tau$. \update{The selection of the form} can have a different basis, depending on the condition, namely, either to complete the backward integration until the perturber is not in the system or, if it is not possible, until some specific time in the past (e.g. until particles that are currently unbound to the perturber were still in its Roche limit). 

\section{Simulations}\label{sec:simulation}
\label{sec:simulations_description}

To test our dynamical friction framework in realistic conditions, such as the presence of an evolving halo, we built two galaxy N-body simulations using the GADGET-4 tree code~\citep{Springel_2021}. In one simulation, a galaxy with no perturber acquires the ${\bf x}_0$ coordinates, and in the other, a galaxy with a perturber obtains perturbed ${\bf x}_1$ coordinates. The initial positions of all particles in both galaxies are identical. We estimated the dynamical friction by comparing the acceleration field of the host particles between the unperturbed and perturbed simulations; see Eq.~\eqref{eq:acc_DF}. We denote \update{this version of the dynamical friction values} as 'actual' onward.

\subsection{General considerations and initial conditions}

We selected a simple model for the simulations so that most effects are due to dynamical friction and not from simulation or model specifics. For the host galaxy, we selected an isothermal sphere model as the observed dynamical mass of galaxies that can be described by isothermal sphere-like distributions \citep{overview_strong_lens_paper}. The density decreases for the isothermal sphere with the radius as $\rho\propto r^{-2}$, and all the locations have the same velocity dispersion $\sigma$. The relationship between the enclosed mass within radius $r$ and the velocity dispersion is $M(<r) = 2\sigma^2r/G$. The total mass of the host is $10^{11}\,{\rm M_\odot}$, distributed in the form of an isothermal sphere within $50\,{\rm kpc}$ (the cut-off radius) with $4\times10^6$ particles. Hence, the mass of a particle is $2.5\times10^4\,{\rm M_\odot}$. 

These simulations were aimed at studying the dynamical friction behaviour in various regions, including sparsely populated outskirts. To cope with these regions and verify the method with minimally noisy data (see Sect.~\ref{sec:tidaleffects} for a discussion on small-scale fluctuations), we picked a sizeable softening length, $b_{\rm soft} = 1\,{\rm kpc}$, to suppress small-scale fluctuations. The softening is described in \citet{Springel_2021}. The parameters controlling the simulation's accuracy were integration accuracy  ($0.0012$), maximum time step ($10\Myr$), and error tolerance to the acceleration ($0.005$). To see the practical accuracy of the test, we re-ran the simulation backwards, from the end to the beginning, and \update{calculated} the shifts and velocity differences between the initial and integrated \update{cases. } For $8\Gyr$, the \update{shift was }$\approx 1.5\kpc$ and \update{the velocity difference was} $\approx 1.5\kms$. As most of our orbit integration is about $0.3\Gyr$, we consider this accuracy acceptable.

\begin{figure}
    \centering
    \includegraphics{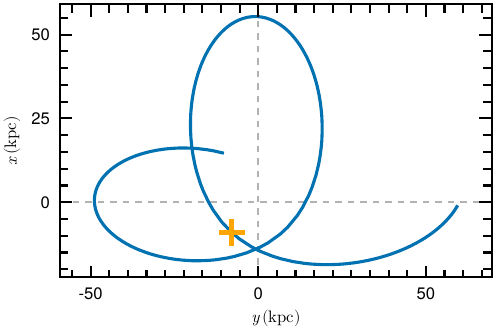}
    \caption{Orbit for the perturber with mass $M_{\rm pert} = 10^9\msun$. The orange cross shows the point corresponding to the snapshot where the acceleration fields are more thoroughly studied. }
    \label{fig:perturber_orbits}
\end{figure}

Initially, we let the simulation reach equilibrium by running it for $2.5\,{\rm Gyr}$. The model retained its spherical shape.\footnote{The number of particles in $10^\circ$ cones followed the Poisson distribution with an accuracy where the Kolmogorov-Smirnov distance between the distributions was $<0.02$.} Then we reset the clock to $t=0$ and introduced the perturber described by the softening potential in~\citet{Springel_2021} (i.e. the perturber has a point-like density distribution). As mentioned, in both simulated galaxies, the initial conditions for the particles were identical; thus, the only differences between these simulations concern the perturber with the mass $M_{\rm pert} = 10^9\,{\rm M_\odot}$. The initial conditions for the perturber were ${\bf x} = (0, 60, 0)\,{\rm kpc}$ and ${\bf v} = (-40, -20, 0)\,{\rm km~s^{-1}}$. Figure~\ref{fig:perturber_orbits} shows the orbit of the perturber. 

\subsection{Approximating acceleration field}\label{sec:sim_accfield}
As input for our dynamical friction calculations, we used the acceleration field of the simulated host galaxies. For a given simulation snapshot, we approximated the host's acceleration field by following the steps outlined in the next paragraphs.

We first set the centre of the coordinates $(0, 0, 0)$ to the system's highest density peak to account for the reflex motion of the host. We then divided the galaxy into radial bins. The mean radial acceleration in each bin was calculated by averaging over all particle accelerations in the bin. The accelerations originate directly from the GADGET-4 output. Finally, we fit a second-order polynomial to the distribution of radial accelerations in each bin separately. The standard deviation of the distribution of radial accelerations minus the fitted polynomial defines the noise in the radial acceleration field in that particular bin as quantified by $a_\sigma$. 

\begin{figure}
    \centering
    \includegraphics{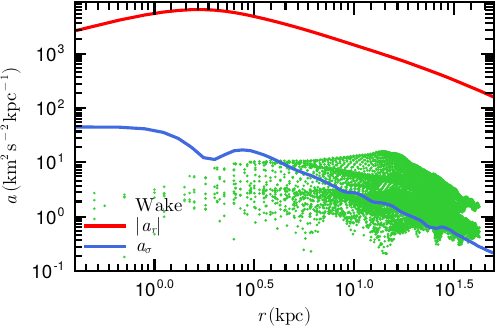}
    \caption{\update{Comparison of different acceleration components in the simulations.} The red line depicts the spherical acceleration field of the unperturbed host's galaxy at $t\approx0.8\Gyr$ (orange cross in Fig.~\ref{fig:perturber_orbits}). The origin of $r$ is the host's centre.  The blue line shows the estimated noise from the host (see text for details), and the green points show the acceleration from various wake points, as in Fig.~\ref{fig:DF_map}. For comparison, the perturber causes an acceleration of $43\accunit$ at $10\kpc$ from \update{the system's} centre. 
    }
    \label{fig:noise_of_acc_host2}
\end{figure}

Through these steps, we derived a smooth, spherical acceleration distribution for the host's potential. Fig.~\ref{fig:noise_of_acc_host2} shows the spherical acceleration field for the unperturbed galaxy as a function of distance to the host's centre. The strength of the estimated noise $a_\sigma$ is approximately two to three orders of magnitude smaller than the average radial acceleration field in the case of the unperturbed galaxy. 

\section{Validation}\label{sec:validation}
The validation was conducted in two parts. \update{First}, we validated the method using the simulations described in the previous section. In this more complex scenario, we quantified the ability of the proposed method to recover the tidal field generated by dynamical friction. \update{Second,} we tested our results against Chandrasekhar dynamical friction \update{in Appendix}~\ref{sec:validation_Ch}.

\subsection{Tidal field origin}\label{sec:validation_simulation}
\begin{figure}
    \centering   
    \includegraphics{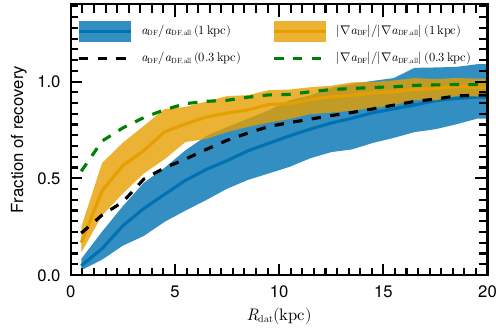}
    \caption{
    Recovery fraction of total dynamical friction or its tides in case of $0.3\kpc$ or $1\kpc$ softening measured by dimensionless acceleration ratios. The simulation was run with $1\kpc$ softening; hence, the contributions from regions dampened by softening would be unresolvable for the simulation. To give an estimate of how much is unresolved, we re-evaluated the dynamical friction using a smaller softening length of $b=0.3\kpc$. Hence, the difference of the lines between different softening values  ($0.3$ and $1.0\kpc$) characterise the contribution of the unresolved part of the dynamical friction.  The band around the lines is formed from the estimates from all the simulation snapshots available. 
    }
    \label{fig:R_dat_global_fraction}
\end{figure}

For practical purposes, we divided the dynamical friction evaluation into three distinct parts based on the distance to the perturber. In this way, the Coulomb logarithm can be split into
\begin{eqnarray}
    \ln\Lambda = \ln{\frac{b_{\rm max} }{ b_{\rm min}}} &=& \ln{\frac{b_{\rm max} }{ R_{\rm dat}} \frac{R_{\rm dat} }{ b_{\rm soft}} \frac{b_{\rm soft} }{ b_{\rm min}} } = \nonumber \\
     &=& \ln \Lambda_{\rm glob} + \ln \Lambda_{\rm loc} + \ln \Lambda_{\rm unres},
\end{eqnarray}
with $b_{\rm max}$ and $b_{\rm min}$ as the size of the system and the minimum impact parameter, respectively, $b_{\rm soft}$ as the softening length, and $R_{\rm dat}$ as the radius of a sphere centred on the perturber containing the particles used to calculate the dynamical friction. In practical calculations, we added a$5\kpc$ edge to $R_{\rm dat}$ when selecting the sample in order to account for particles leaking in or out of $R_{\rm dat}$, but for the acceleration calculations, we strictly used the data within it. We define $R_{\rm dat}$ as the local region around the perturber:
\begin{eqnarray}
    |{\bf x} - {\bf x}_{\rm p}(t=0)| \le R_{\rm dat} \label{eq:def_Rdat}
\end{eqnarray}
Figure~\ref{fig:R_dat_global_fraction} shows how the different splits capture different amounts of dynamical friction. The fraction of $100\%$ is with $R_{\rm dat}=\infty$ capturing all of dynamical friction.
The acceleration field is affected by the global perturbations induced by the perturber; therefore, $R_{\rm dat}$ must be large in order to obtain an unbiased estimate of the acceleration field. The larger $R_{\rm dat}$, the less \update{computationally} efficient the orbit rewind method will be, where $R_{\rm dat}=\infty$ means effectively reconducting  the entire simulation. For the current case with a $10\kpc$ sized region and $50\kpc$ sized galaxy, the average fraction of stars to evaluate is $0.8\%$. Nonetheless, the tidal field is less prone to global perturbations, and it converges to its total value for smaller $R_{\rm dat}$ than for the acceleration field. For this reason, we assessed the accuracy of the method by recovering the tidal field, and we fixed $R_{\rm dat} = 10\,{\rm kpc}$, which is sufficient for recovering most of the tides of dynamical friction. 
Discarding the global part of the dynamical friction estimation leaves the doubt as to whether a sufficient maximum $R_{\rm dat}$ value is used. From the practical perspective, the question can be answered `on the fly' at every evaluation. Having a set of points ${\bf x}_0$ and ${\bf x}_1$, the $R_{\rm dat}$ dependence within some maximum value can be constructed. If the dependence reaches some plateau (as in Fig.~\ref{fig:R_dat_global_fraction}), then the chosen $R_{\rm dat}$ is sufficient. 

To quantify the recovery of the tidal field at the perturber's position due to dynamical friction, we compared the fields constructed using the perturbed and unperturbed states. \update{The unperturbed states were} obtained directly from the simulation, \update{while} the perturbed positions are from the orbit-rewinding of the unperturbed points \update{or from the perturbed simulation}. 

\subsection{Reconstruction properties}\label{sec:reconstr_prop}

The recovery of tidal dynamical friction assumes some choices in the implementation. We used a particle-based description with Eqs.~\eqref{eq:anp} and \eqref{eq:ap}, and we assumed the kernel to be a Plummer sphere (see Eq.~\eqref{eq:Plummer_profile}) with $b=1\kpc$ to account for softening. The orbit integration was performed using the \update{Runge-Kutta-Felberg method} with an adaptive step size while keeping the velocity accuracy below $10^{-4}\kms$. The $N$ parameter that controls the extent of orbit integration from Eq.~\eqref{eq:N_definition} is $N=10$ in our fiducial case. The step size for the numerical derivative (see Eq.~\ref{eq:theory_tidal})  is $\Delta = 2\kpc$. We note that it should be larger than the softening radius and similar to the radii around the perturber at which the lopsidedness is to be evaluated. The previous values were varied in order to observe the sensitivity of the recovery. The total number of iterations of the method (see Sect.~\ref{sec:basic_framework}) was one, and we did not have to approximate the local dynamical friction estimate with Eq.~\eqref{eq:adf} but instead used the sum directly. 

\begin{figure}
    \centering
    \includegraphics{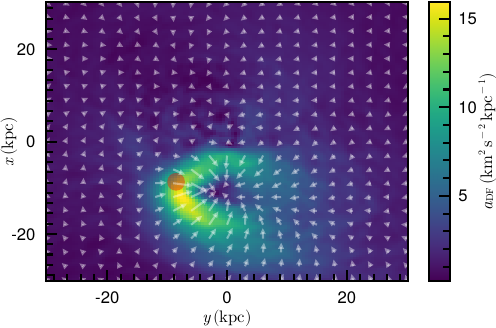}
    \caption{Acceleration field from the dynamical friction at $0.8\Gyr$ from the beginning of the simulation. The point in the orbit is shown in Fig.~\ref{fig:perturber_orbits}. In the current example, the wake is about $10\kpc$ behind the perturber and behaving as a sink for ${\bf a}_{\rm DF}$. }
    \label{fig:DF_map}
\end{figure}

\begin{figure*} 
    \includegraphics{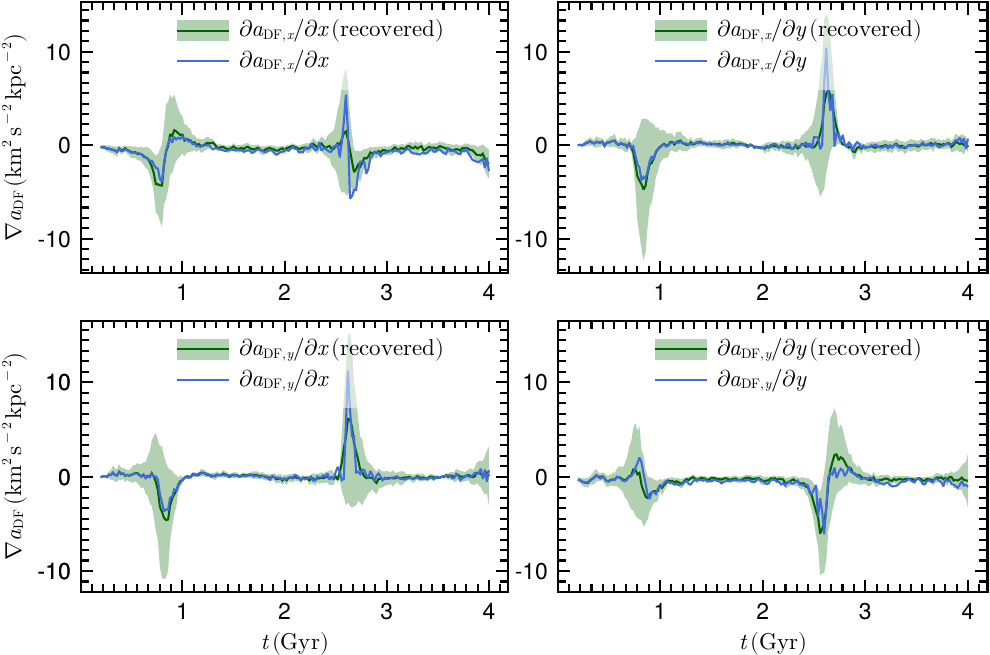}
    \caption{Recovery of the tides from dynamical friction for the fiducial case (see Table~\ref{tab:residuals} for numeric characterisation of the recovery). Each panel shows one component of the tides. The ratio of residuals and uncertainty are shown in Fig.~\ref{fig:residual_TDF}. The uncertainty (green corridor) was evaluated from the summing in Eqs.~\eqref{eq:ap} and \eqref{eq:anp} in the recovery case (green line). We note that the uncertainty is not normally distributed. The blue line shows the actual value.
    }
    \label{fig:validation_TDF}
\end{figure*}

The acceleration field due to dynamical friction at 0.8 Gyr (orange cross in Fig.~\ref{fig:perturber_orbits}) is shown in Fig.~\ref{fig:DF_map}. As the perturber orbits in the $x$-$y$ plane, we only took interest in the  $x$ and $y$ components, namely, $\partial a_{{\rm DF},x}/\partial x$, $\partial a_{{\rm DF},x}/\partial y$, $\partial a_{{\rm DF},y}/\partial x$, and $\partial a_{{\rm DF},y}/\partial y$. The comparison of the actual and recovered derivatives as a function of time are shown in Fig.~\ref{fig:validation_TDF}.

\begin{figure}
    \centering\includegraphics{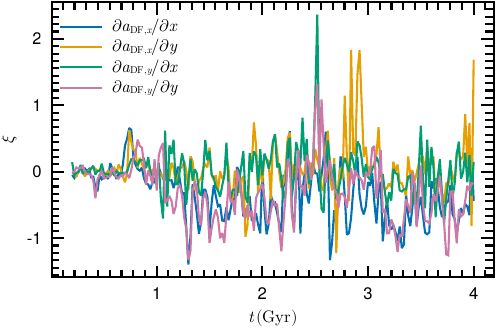}
    \caption{Evolution of the recovery parameter defined with Eq.~\eqref{eq:residual_def}. In other words, this figure shows the residuals of Fig.~\ref{fig:validation_TDF} divided by the uncertainty. This figure complements Table~\ref{tab:residuals}. }
    \label{fig:residual_TDF}
\end{figure}

For numerical estimates of the recovery, we used the following estimator 
\begin{equation}
\xi =(\nabla {\bf a}_{\rm DF }^{\rm actual} - \nabla{\bf a}_{\rm DF}^{\rm recovered} ) / \sigma_{\nabla {\rm a}_{\rm DF}}.\label{eq:residual_def}
\end{equation}
Estimates were calculated at each simulation snapshot for each of the derivatives considered. In the equation, $\sigma_{\nabla {\rm a}_{\rm DF}}$ denotes the uncertainty of $\nabla{\bf a}_{\rm DF}^{\rm recovered}$ and was evaluated using the bootstrap method \citep{bootstrap_mainpaper}. We did not use the relative error, as the values oscillate around zero, and divisions by small values do not describe the actual recovery. The evolution of $\xi$ values is illustrated in Fig.~\ref{fig:residual_TDF}.

Figure~\ref{fig:residual_TDF} shows $\xi$ values as a function of time for the considered derivatives. In addition, Table~\ref{tab:residuals} summarises the value of $\xi$ averaged over time and its components. We note that in the case of perfect recovery, the mean of $\xi$ is zero, and the standard deviation of the values is minimal. Standard normal distribution has a standard deviation of one, but we emphasise that due to the existence of high leverage contributors in the sum in Eqs.~\eqref{eq:anp} and \eqref{eq:ap}, the normal assumption is not valid. 

We concluded that the current method is able to recover the tides from the wake of dynamical friction. The first stages (before first pericentre passage) have a better quality of recovery, which is expected, as the system remains in a simpler state. 

\subsection{Verification of the adiabatic introduction}\label{sec:verification_of_adiabatic_introduction}
In Sect.~\ref{sec:adiabatic_introduction_theory}, we pointed out that for some stars, the condition for the extent of the orbit integration cannot be achieved and introduced an alternative to it. In this section, we test this alternative:  the adiabatic introduction.

We used the sigmoid function for the $\tau$ 
\begin{eqnarray}
    \tau = \frac{1}{1+\exp[(t-t_0)/\Delta\! t]}, \label{eq:ai_implementaion}
\end{eqnarray}
which has two plateaus and a smooth transition region between them. In our implementation, the first plateau corresponds to the time when the perturber was not in the system, and the second plateau corresponds to when asymptotic current perturber mass was achieved. This mimics the buildup and provides a well-defined orbit integration end condition and a constant perturber mass region in order to get the highest accuracy of the part of the wake closest to the perturber, which presumably has the highest impact on the tidal dynamical friction and forms last. In Eq.~\eqref{eq:ai_implementaion}, the $t_0 = 0.35\Gyr$ denotes the mid-point of introducing the perturber, and $\Delta\!t=50\Myr$ is the buildup speed. 

\begin{figure}
    \centering
    \includegraphics[width = \columnwidth]{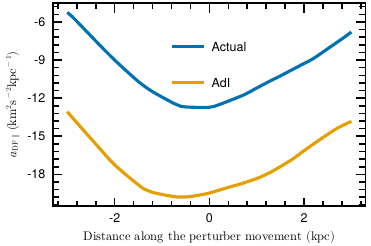}
    \caption{Dynamical friction as a function of the distance to the perturber along the movement vector of the perturber. The blue line denotes the actual value (see Sect.~\ref{sec:reconstr_prop}), and the golden line (denoted as AdI) shows the value recovered using the adiabatic introduction. The main aspect of this result is that the total values of the dynamical friction do not match as expected from the absence of the global wake. However, the slopes match, indicating that the tides from the dynamical friction are robust in the current case. }
    \label{fig:AI_slopes}
\end{figure}

The two flavours of the recovery method of the (tidal) dynamical friction are similar. In Fig.~\ref{fig:AI_slopes}, we present the dynamical friction values along the path traced by the perturber. The dynamical friction vector is projected to the same \update{trajectory} line. The recovered value does not match with the actual one, having an about $30\%$ mismatch and indicating that the global effects influence recovery. However, the slopes of the dynamical friction match well, indicating the tidal field that has a local origin (see Sect.~\ref{sec:tidaleffects}) is well recovered. 

Overall, we conclude that the adiabatic introduction recovers the (tidal) dynamical friction as well as the 
Eq.~\eqref{eq:practical_main_assumption}-based \update{version}, but it has both advantages and disadvantages. It has a wider application range and can be used without knowing the noise levels of the galaxy $a_\sigma$ (in such a case, $t_{\rm far}$ is shared between all particles). Regarding disadvantages, the integration of orbits is no longer minimal, causing an accumulation of $x_{\rm mismatch}$ and possibly biasing the results further. The longer orbit integrations also increase calculation time  (by about an order of magnitude for the current example).

\begin{table}
\caption{Table quantifying the recovery of tides of dynamical friction. It shows the dependence on the step size of derivative ($\Delta$), the softening length ($b$), and the extent of integration, $N$; see Eq.~\eqref{eq:N_definition}. The definition of $\xi$ is given with  Eq.~\eqref{eq:residual_def}; the terms $\langle \xi \rangle$ and $\sigma_\xi$ denote the mean and standard deviation of $\xi$ time evolution, respectively. The table complements Fig.~\ref{fig:residual_TDF}.  }
    \label{tab:residuals}
    \centering
    \begin{tabular}{lll}
    \hline
 & $\langle \xi \rangle $ & $\sigma_\xi$ \\
\hline
Fiducial & -0.149 & 0.443 \\
$\Delta = 1.0\kpc$ & -0.102 & 0.341 \\
$\Delta = 3.0\kpc$ & -0.168 & 0.525 \\ 
$b = 0.5\kpc$ & -0.171 & 0.555 \\ 
$b = 1.5\kpc$ & -0.131 & 0.405 \\
$N = 5$ & -0.151 & 0.433 \\
$N = 20$ & -0.144 & 0.442 \\
\hline
    \end{tabular}
\end{table}

\section{Sampling noise}
\label{sec:sampling_noise}
As shown in Sect.~\ref{sec:tidaleffects}, in dynamical friction estimates, one also needs to be aware of the sampling noise. As the previous section showed, we were able to recover the local wake of dynamical friction adequately; hence, the current method is able to study how the local sampling noise affects dynamical friction. In this section, we study noise only from a physical basis without any attempt to compensate it with softening, that is, the kernel in Eqs.~\eqref{eq:anp} and \eqref{eq:ap} is the \update{kernel of the} point source. 

If we have a region where we want to evaluate the dynamical friction as accurately as possible, a problem arises when we do not observe all the stars, as the dynamical friction evaluation then has some uncertainty and inaccuracy. Even if the overall particle distribution is statistically identical, realisations of this distribution vary, thus varying the estimates of the dynamical friction. 

From the simulated galaxy described in Sec.~\ref{sec:simulations_description}, we created different realisations of a sphere with radius $R_{\rm dat} = 2\kpc$ centred on the perturber at $t\approx0.8\,\rm Gyr$ (orange cross in Fig.~\ref{fig:perturber_orbits}) and studied the influence of the sampling noise. This region contains $N_0 = 1943$ particles, corresponding to a mass density of $0.0014\msun~{\rm pc^{-3}}$ (for comparison, the mass density in the solar neighbourhood is $0.097\msun~{\rm pc^{-3}}$ \citep{McKee:2015}, or it is the same value as the galactic DM density $24\kpc$ \citep{McMillan17}), and has a total mass of $M_{\rm tot} = 5\times10^{7}\msun$. The statistical analogues share these parameters and are calculated by sampling from the phase-space distribution function of the simulated particles in that region. Specifically, by placing the perturber at the centre of the reference frame and using spherical coordinates, we first sampled the radial number density and then the angles from the perturber. At the obtained position, we sampled from the constructed radial and tangential velocity distribution functions. Hence, we generated statistical analogues of the region of interest. For each analogue $i$, we calculated the local acceleration field due to the dynamical friction at the perturber's position (i.e. $a_{\rm DF}^i$) to where we added the global contribution from the simulation data. To assess the dependence of the dynamical friction estimate on the number density of the particles, we combined realisations in the following manner. We combined $N_{\rm stack}$ random realisations and estimated the acceleration field as 
\begin{eqnarray}
    a_{\rm stack} = \frac1{N_{\rm stack}} \sum_i a_{\rm DF}^i.
\end{eqnarray}
In this way, the effective particle mass for the stacked set of particles is
\begin{eqnarray}
    m_{\rm particle, eff.} = \frac{M_{\rm tot}}{N_0 N_{\rm stack}}.\label{eq:massparticle}
\end{eqnarray}
As we were summing in $a_{\rm stack}$, we were able to evaluate its uncertainty by stacking different analogues, similar to bootstrapping. If the $m_{\rm particle}$ reached the physical average of the stellar initial mass function (IMF), we would reach the noise levels the physical dynamical friction can have.

\begin{figure}
    \centering
    \includegraphics{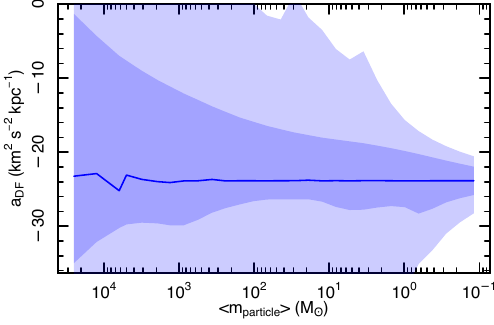}
    \caption{Dependence on the physical fluctuations in the dynamical friction due to the discreteness the of stars. The $a_{\rm DF}$ describes the estimate of the dynamical friction by combining different analogues of the region. The dark- and light-blue bands represent the $1\sigma$ and $2\sigma$ uncertainty of the $a_{\rm DF}$. See Section~\ref{sec:sampling_noise} for further details. }
    \label{fig:physical_limits}
\end{figure}

Figure~\ref{fig:physical_limits} shows the mean of this distribution of $a^i_{\rm DF}$ values with the corresponding uncertainty as a function of the \update{particle mass in a fixed physical density}. From the figure, we observed that the sampling noise increases with a decreasing number density (larger mass).  We note that for some setups, the dynamical friction is physically indeterminable. At a smaller particle mass, the sampling noise introduces a 10\% uncertainty at the 1$\sigma$ level. In case of solar mass particles, the uncertainty is about 23\%. 

We note that the described stacking procedure has a physical limitation: The mass of a single particle cannot fall below the average mass of a star of a brown dwarf. Therefore, Eq.~\eqref{eq:massparticle} poses a strict limit on how many realisations we can combine or stack. A similar restriction should be posed to numerical integration over phase-space density:
\begin{eqnarray}
    f({\bf x}, {\bf v}){\rm d}{\bf x}~{\rm d}{\bf v}\gtrsim \langle m_{\rm particle}\rangle.\label{eq:minimum_phase_space_element}
\end{eqnarray}
The $\langle m_{\rm particle}\rangle$ denotes the physical mass of an average star (or DM particle in case of halo dynamical friction inference). Suppressing the noise further is not \update{physically justified. }
Analogously, the spatial element of adaptive integration cannot be smaller than \update{the element containing the mass of the smallest stars} 
in order to keep the integration accuracy compatible with physics.

\section{Discussion}\label{sec:discussion}
This paper has two aims: to introduce a method to calculate local dynamical friction that is usable for perturber distortion evaluation and to show that dynamical friction has intrinsic fluctuations from the sampling noise. As with all dynamical friction methods, the current one has its limitations but also regions where it excels. Method limitations originate mainly from using only the positions of particles in the vicinity of the perturber (within $R_{\rm dat}$). Having this limitation, we are only able to include contributors to dynamical friction that are inside of the perturber and not the effects of a global wake. To summarise, the limitations and advantages are following.

    The method can capture slightly more than half of the total drag force arising from dynamical friction from a finite $R_{\rm dat}$ (see Sect.~\ref{sec:validation_simulation} and Fig.~\ref{fig:R_dat_global_fraction}), but at the same time, the method provides input to recover the tidal distortions of the perturber due to the wake of dynamical friction. It assumes the unresolved softening part produces no contributions. Upon a significant suppression of noise from the softening, we recover a larger fraction of the total (tidal) dynamical friction. 
    
    The method does not hold any assumption about the host's symmetry, stationarity, or equilibrium. Hence, the dynamical friction effects can be inferred in very complex environments, and thus the practical applications are limited only by our ability to describe the system fully, including observational limitations. This method can cope with systems without simplifications. For example, if the velocity distribution is not Gaussian but is described as Gaussian, dynamical friction misevaluation alters infall times (and hence dynamical friction) by a factor of $50\%-300\%$ \citep{Just:2011}, or \update{the absence of} the long-term tidal effects \update{alter the friction} by about $10\%$ \citep{Roshan_2022}. The tidal effects caused by dynamical friction are not a well-studied topic; hence, we are unable to compare the tidal contributions directly, and we are satisfied with the total dynamical friction. It pushes the limit to evaluate the tidal dynamical friction to cases where complexity is very high, for example, where a pair of perturbers, such as the Large and Small Magellanic Cloud, produce wakes and tides.
    
    We showed that the noise in dynamical friction originates predominantly from the vicinity of the perturber, Sect.~\ref{sec:tidaleffects} and Fig~\ref{fig:R_dat_global_fraction}, 
    arising from statistical fluctuations of the density field due to a finite number of stars. The current modelling enables us to quantify this effect.

The current method relies heavily on an orbit integration between the past and present \update{time}, and one can ask if it would not be simpler to only make two simulations in cases where one wishes to evaluate dynamical friction. For studies that are aimed at examining the behaviour of dynamical friction in various setups, it is a valid approach. But if one wishes to infer the physical aspects of some specific host or perturber, this does not suffice. The main reason is that the initial conditions corresponding to the observations are not known. 

The current method is applicable to several studies, with the highest impact potentially on problems that involve the evaluation of tidal dynamical friction in a time- or location-dependent setup. 
\begin{itemize}
    \item {Lopsided galaxies:} \citet{Kipper_lopsided} has demonstrated that in some cases, a galaxy moving through a medium of uniform density can produce a wake behind it (consisting of unbound particles) that both slow down the galaxy and cause a tidal field, producing lopsidedness in the galaxy. Due to its ability to describe complex situations, the current method allows for diversification of the studies of environments where the lopsidedness can form. 
    \item {Simulation recipes:} Upon calibrating (using machine learning methods or otherwise) the dependence of precise local dynamical friction on host environmental parameters and incorporating such information into simulations, the calibration will lead to an improvement of the resolution of processes involving the sinking of super massive black holes, or their seeds  \citep{seedsdontsink}, to the centres of galaxies.
    \item {Heating of globular cluster outskirts:} 
    Changes in gravitational potential, either through fluctuations in dynamical friction or the change of the location of the wake with respect to the globular cluster,  cause  heating in stellar systems. The wake is one example of a time-dependent potential. Upon resolving the time dependency of the gravitational potential in the vicinity of globular clusters, the heating can be estimated. The current method allows for the wake-induced potential changes to be evaluated. For example, \citet{Wan_2021} showed that processes included in their simulation cannot explain the velocities in the globular cluster outskirts, but the current method can provide some additional possibilities for the explanation.
    \item {Magellanic Clouds}: A lot of interest has emerged in studies of the Large Magellanic Cloud due to the possibility that its mass is significantly larger than previous studies have indicated \citep{Erkal:2021}. 
    Also, \citet{Cullinane_2022} reported the perturbed state of the Magellanic Cloud outskirts.  If the dynamical friction is capable of causing the Magellanic Clouds to sink into the Milky Way, it is reasonable to speculate that the wake can alter its outskirts. The current methodology is able to cope with multiple perturbers and evaluate the contribution of wakes to the shape distortions of the perturbers. 
    A similar methodology has been used by \citet{CorreaMagnus:2022} to estimate the mass of our Galaxy. 
    \item {Tail of Ram-pressure stripping:} Galaxies moving through the intergalactic medium are affected by ram-pressure stripping. A jellyfish galaxy tail should coincide with the wake of dynamical friction. The tidal wake caused by the dynamical friction could lead the streams and tails of ejected material to an overdensity that follows the acceleration field     \citet{Boselli22}, \citet{Triantafyllaki}. Previous works using N-body-SPH simulations focusing on the ram-pressure scenario \citet{Mastropietro09} on the Large Magellanic Cloud have tried to explain observations of HI regions and recent star-formation events. Using this method, it is possible to test whether this star formation can be localised in a region where the streams are tugged by the acceleration from the wake. 
    \item {DM inference:} Once it is possible to evaluate the dynamical friction from the wake's tidal field of lopsided galaxies, the formation of lopsidedness can be studied in the context of different DM scenarios. Various DM models show differences in the wake \citep{Hui_2017} of the DM; hence, the current method has a potential to distinguish among them based on the effects they cause on the lopsided galaxy. 
    \item {Open cluster (OC) stream asymmetry:} \citet{Pflamm23} investigated the asymmetric evaporation of stars in the leading and trailing tidal arms from four open star clusters. With this method, we plan to measure if this asymmetry could be a result of a difference in the dynamical friction effects in the tidal arms. The tidal radius is at most a few hundred parsecs from the clusters, depending on the location \update{and the velocity of the OC}. The wake can be at similar distances but not necessarily aligned with the galactic tidal field and should have an overlap with the trailing arm of the stream. In addition, the location of the Lagrangian points becomes altered compared to the isolated Milky Way-OC system, affecting the formation and shape of the stream.
\end{itemize}

To use the proposed method in real galaxies, one of the concerns is the selection of stars. Namely, this pertains to how to separate the stars belonging to the perturber (or the ones that have just left the perturber via tidal stripping) from the host stars. The strategies for coping with this problem are two-fold, depending on if the individual stars are resolved or not. If the stars are resolved, such as in the case of the Milky Way halo, the split can be done by using either kinematics (co-moving host stars typically have larger velocities with respect to the perturber than the escape velocity from the perturber) or detailed chemical composition (metallicities of the perturber stars are, as a rule, in a narrow range). For example, \citet{Battaglia:2012} found the recipe to separate the interlopers using specific spectral lines. If the single stars of a galaxy are not resolved, the only possible approach is to compare the stellar population properties of the host and the perturber. For example, the J-PAS survey \citet{JPAS} provides an excellent set of filters to obtain the basis for the population separation.  

\section{Summary and conclusions}
\label{sec:conclusion}
We developed a flexible semi-analytical method to  calculate tidal dynamical friction of a perturber in a complex matter distribution. The current method reformulates dynamical friction as the difference between the acceleration fields of the stars with and without the perturber. Hence, we need two acceleration fields, or particle configurations, to infer dynamical friction. One configuration comes from observations, and the second has to be computed. In Sect.~\ref{sec:validation} we validated the method.

The central assumption of the proposed method is that for each particle, there was a time in the past ($t_{\rm far}<0$) when the influence of the perturber was insignificant compared to the host. The $t_{\rm far}$ is obtained by integrating the particle's and perturber's orbits towards the past and continuously checking the validity of the condition \eqref{eq:practical_main_assumption}. Once achieved, the particle coordinates at $t_{\rm far}$ form the initial conditions for calculating the two configurations. The perturbed state corresponds to integrating back to the present in the presence of the perturber potential, and the non-perturbed \update{state corresponds to } the absence \update{of the perturber}.\footnote{Practical calculations require calculating back to the present only once, as the other configuration is the observation.} After repeating this procedure for all the particles, we obtained the basis for calculating the acceleration field and hence the dynamical friction. 

The main conclusions of this study are as follows:
\begin{itemize}
\item We designed a method that allows for accurate local dynamical friction estimates in a system with various complexity levels where the phase space density is describable either analytically or as a set of particles. 
\item We successfully recovered the Chandrashekhar dynamical friction in a homogeneous and continuous medium (as shown in Fig.~\ref{fig:Ch_all}).
\item A drastically different base-assumption set (see Sect.~\ref{sec:basic_framework}) allowed us to calculate local dynamical friction in a broad range of possible environments and configurations (see Sect.~\ref{sec:discussion}), such as infalling systems. 
\item The method is limited to evaluating the local aspects of dynamical friction that are the sources of tides from the wake of dynamical friction. These tides lead to shape distortions of the perturber. 
\item Our assumption set and dynamical friction formulation suggest that the points contributing to dynamical friction have an $r^{-3}$ distance dependence from the perturber, which makes dynamical friction more prone to sampling noise than total acceleration (see Sect.~\ref{sec:tidaleffects}). The current method is able to distinguish the issues arising from the sampling noise. 
\item The uncertainty from sampling depends on the number density of particles. In Sect.~\ref{sec:sampling_noise}, we showed  the sampling noise associated with dynamical friction for various number densities. The fluctuations from the sampling noise do not reach insignificant levels with an increasing number density, showing that the fluctuations are inherent in the nature of the dynamical friction of point sources. The level of noise for our setup with density $0.0014\msun\,{\rm pc^{-3}}$ has an intrinsic scatter of $10\%$ for dynamical friction (Sect.~\ref{sec:sampling_noise}). 
\end{itemize}

 Although there are many possible applications of the presented method for computing dynamical friction, this paper covers only one: the intrinsic noise of dynamical friction. We showed that the uncertainty (fluctuations) in dynamical friction calculations is related to the number density of particles. In reality, the number density of stars is not sufficient to firmly 
 calculate dynamical friction. For example, when assuming that a stellar halo consists \update{of stars with the mass of an} of an average brown dwarf, there is always a  $\sim10\%$ uncertainty in dynamical friction (see Fig.~\ref{fig:physical_limits}).

\section*{Acknowledgements}
We are grateful to referee for useful recommendations and an idea which has immensely improved the paper. The present study was supported by the ETAG projects PRG1006, PSG700, and by the European Structural Funds grant for the Centre of Excellence "The Dark Side of the Universe" (TK133). We applied in this study R statistical environment \citep{ig96}, C++ and Julia \citep{bezanson2017julia, DanischKrumbiegel2021}.

\bibliographystyle{aa}

\begin{appendix}
\section{Chandrasekhar dynamical friction}\label{sec:validation_Ch}
We validated our method against the classical Chandrashekhar formulation under the assumptions of an infinite, homogeneous, and isotropic environment. We picked the Chandrasekhar dynamical friction for comparison, as it is an intrinsically local method without any global modes, which our method is capable of reproducing asymptotically. Reproducing Chandrasekhar results allowed us to validate the method  in the absence of global modes. 

We adopted a setup that mimics a globular cluster in the solar neighbourhood: a perturber with mass $M=2\times 10^5\,{\rm M_\odot}$ is moving with a velocity $v=30\kms$ relative to the solar neighbourhood in an environment with velocity dispersion $\sigma = 20\kms$ and density $\rho = 0.1\,{\rm M_\odot}{\rm pc}^{-3}$. Thus, the perturber is almost co-moving with the rotation of the disc. We characterised the perturber with a Plummer profile
\begin{eqnarray}
    \Phi = \frac{-GM}{\sqrt{r^2 + c^2}},\label{eq:Plummer_profile}
\end{eqnarray}
with $c=0.04\kpc$. 

The Chandrasekhar dynamical friction equation assumes that a point-massed perturber with mass $M$ and velocity ${\bf v}$ moves in an infinite, uniform, and isotropic collisionless background. Under these assumptions, the acceleration of the perturber can be calculated as
\begin{eqnarray}
    \frac{{\rm d}{\bf v}}{{\rm d}t} = -\frac{4\pi G^2\rho M \ln\Lambda}{v^3}\left[\text{erf}\left(
    \frac{v}{\sqrt{2\sigma^2}}
    \right) - 
    \sqrt{\frac{2}{\pi}}
    \frac{v}{\sigma}
    e^{-0.5v^2/\sigma^2}\right]{\bf v}\label{eq:DF_Ch},
\end{eqnarray}
where the only unknown parameter of the Chandrasekhar equation is the Coulomb logarithm $\ln\Lambda$. The slight difference arising from our use of the Plummer profile instead of a point mass is negligible.

We calculated the dynamical friction as outlined in Sect.~\ref{sec:method} using a continuous phase-space description of the system. The orbit integration was evaluated using a Runge-Kutta-Fehlberg adaptive step approach. To evaluate Equations \eqref{eq:analy_nonpert} and \eqref{eq:analy_pert} numerically, the integrals had to converge, but they are divergent, just as in the Chandrasekhar case. Therefore, we integrated up to a distance $r_{\infty} = |{\bf x}(t_{\rm far}) - {\bf x}_{\rm p}(t_{\rm far})| = 4\kpc$ from the perturber, as at the distance $r_{\infty}$ from the perturber, we consider conditions \eqref{eq:main_assumption} and \eqref{eq:practical_main_assumption} to be fulfilled. At that distance, the acceleration from the perturber is about $0.05\accunit$; for comparison, the acceleration towards the Galactic centre in the solar neighbourhood is $\approx6400\accunit$ \citep{Kipper_OAM_nonstat}. 

\begin{figure*}
\includegraphics[width=\textwidth]{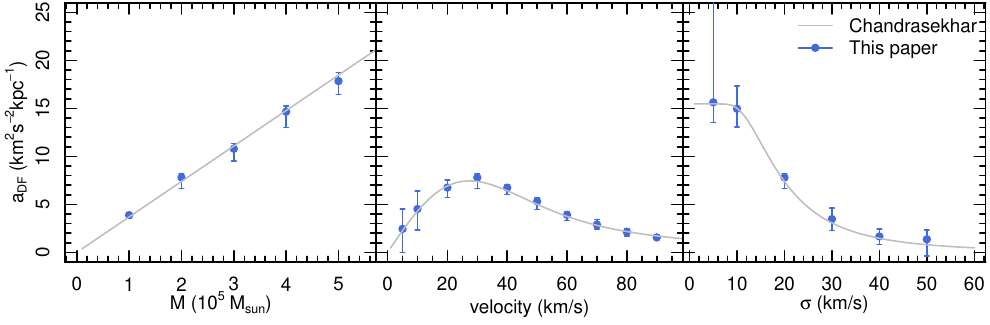}
\caption{ Dynamical friction in an infinite, homogeneous, and isotropic medium calculated from the Chandrashekhar method (grey solid line) and our method (blue circles). The vertical bars in all the panels depict the dynamical friction changes in the vicinity of $50\pc$ from the perturber. We note that they do not describe error or uncertainty. 
\update{The dynamical friction dependency on the mass (left panel), velocity (central panel), and velocity dispersion (right panel) is recovered. }
}
\label{fig:Ch_all}
\end{figure*}

The Chandrasekhar dynamical friction contains an undetermined constant that characterises the extent of the regions in which the dynamical friction is evaluated. As the constant is unknown, the direct values are not comparable; hence, we relied on the recovery of the dependency variables $M$, $v$, and $\sigma $ for validation. Figure \ref{fig:Ch_all} shows a comparison between the estimates of the dynamical friction from the current method and from the Chandrashekhar method when varying the perturber's mass $\rm M$ (left panel), its velocity (middle panel), and the velocity dispersion of the particles in the background medium (right panel). In our calculations, we assumed the Coulomb logarithm is $\ln\Lambda = 0.65\ln (r_\infty / c)$ in order to match the normalisation of the Chandrasekhar method to ours. This figure shows, as indicated by the vertical bars, the range of acceleration values within $50\pc$ around the perturber.  From the figure, we observed that the local dynamical friction values are consistent for both methods, and the trends remain robust. In conclusion, the current method successfully recovers the Chandrashekhar dynamical friction in the case of an infinite, homogeneous, and isotropic medium; hence, the local wake is reproduced. 
\end{appendix}
\label{lastpage}
\end{document}